\documentclass[twocolumn]{aastex631}
\usepackage{graphicx} 
\usepackage{amsmath}
\usepackage{mathtools}
\usepackage{enumitem}
\usepackage{tikz}
\usepackage{soul}
\usepackage[normalem]{ulem}
\usepackage{CJK}
\usepackage{hyperref}
\hypersetup{
    colorlinks=true,
    linkcolor=blue,
    filecolor=magenta,      
    urlcolor=blue,
    pdftitle={Overleaf Example},
    pdfpagemode=FullScreen,
    }

\graphicspath{{./}{figures/}}
\shorttitle{Lit or cringe?}
\shortauthors{铁蛋}
            
\begin{document}
\begin{CJK*}{UTF8}{gbsn}

\title{As a matter of colon: I am NOT digging cheeky titles (\textit{no, but actually yes $:>$})}

\author{Joanne Tan}
\altaffiliation{The authors are responsible for the contents of this paper, which do not in any way represent the views of their employers. Their views are their own, no matter if scrutinized through a microscope or a telescope.}
\affiliation{Earth, Maybe. Perhaps an Extragalactic Existence - Essentially Humanoid.}

\author{Tie Sien Suk}
\affiliation{34.413977, -119.842951}

\begin{abstract}
What's in a name, a poet once asked. To which we present this work, where we investigate the importance of a paper title in ensuring its best outcome. We queried astronomy papers using NASA ADS and ranked 6000 of them in terms of cheekiness level. We investigate the correlation between citation counts and (i) the presence of a colon, and (ii) cheekiness ranking. We conclude that colon matters in the anatomy of a paper title. So does trying to be cheeky, but we find that too much cheekiness can lead to cringefests. Striking the right balance is therefore crucial. May we recommend aiming for a level 4 cheekiness on a scale of $1-5$. 
\end{abstract}

\section{Introduction}
\label{sec:1-intro}

Publications, and the ensuing citations, are deemed \textit{THE} currency for academics. As such, we try every possible way to attract people to our papers amidst the deluge of 60+ papers that are pushed to arXiv each day. Some of these efforts include: 
\begin{enumerate}[label=(\roman*)]\itemsep -4pt
    \item submitting the moment the submission window opens using an automated submission script (you know who you are, and which script I'm talking about $--$ or in zoomer terms, \textit{iykyk}), such that our paper appears first on the listing. There seems to be a paradigm that the lower your paper appears on the listing, the more likely it will get scrolled past and forever be buried in the ether;
    \item attaching a Youtube video along with our submission, thereby condensing our years of work into 5-min videos, which does not help in reducing the sense of existentialist crisis in academics (well, maybe just me); 
    \item posting a long thread to the bird app announcing our very special ``paper day!" and explaining our science (in hopes we get dozens of retweet and quote-retweet), and many more.
\end{enumerate}

In addition, another popular way to attract attention to our papers is to title our papers appropriately. We all know that first impression matters. As we ponder what we should name our child/pet/plant/vehicle/telescope/software, so we also ponder how we should craft the best title for our papers. Readers have surely come across a specific strand of paper titles: cheeky \textit{clickbaity} paper titles with various pop culture references mixed in. Perhaps the idea is that embellishing our papers with cheeky titles will make our papers sound more inviting and less formidable, which in turn inspires the reader to continue reading. Or maybe they, by virtue of being more memorable, might help increase our citation counts in the long run. Or perhaps they make the paper sound more (counter-intuitively) credible, which should hopefully also generate more citations. 

No matter the intentions behind the birth of these titles, in this work, we investigate if the cheekiness of a paper title truly has an impact on the number of citations. In \S\ref{sec:2-methods}, we outline our cutting-edge methods to assemble the relevant data needed for this study. We present our results in \S\ref{sec:3-results}, which will hopefully inform fellow researchers on the best strategy to ensure paper longevity. We conclude in \S\ref{sec5: conc}. 

\section{Methods}
\label{sec:2-methods}
We query the SAO/NASA Astrophysics Data System (ADS) via its API\footnote{\href{https://ui.adsabs.harvard.edu/help/api/}{https://ui.adsabs.harvard.edu/help/api/}} for papers from 2000 to 2023, provided the paper:
\begin{enumerate}[label=(\roman*)] \itemsep -4pt
    \item is refereed,
    \item is of type article, 
    \item has no more than $4^2$ authors (to weed out large collaborations who tend to self-cite), and
    \item is published in ApJ, ApJS, MNRAS, AJ, A\&A, A\&ARv, JCos, or JCAP.
\end{enumerate}

Regarding criteria 4, we tried to restrict the query to the astronomy database, but found this to not be restrictive enough, as the results contain papers from geology to oceanography (perhaps yet again another attempt by academics to cast a wide net on paper submission...?). Our query resulted in 187,659 total papers. Assuming a speed reader takes 2 seconds to read a paper title, they will need about 104 hours to just read ALL of the titles. This immediately tells you how yuge our paper arXiv is!

Next, we tackle the issue of hunting for cheeky paper titles. Over years of observations, we have come to notice that cheeky titles usually contain colons, ``:''. Therefore, we use the presence of a colon in a title as a proxy for cheekiness. Out of 187,659 papers, we filtered out 38,544 papers (20.5\%) that contain a colon in their titles. We then manually rank\footnote{ML enthusiasts might wonder why we don't train a model or two, but due to limited time, limited training sets, and the complexity of our specific sentiment analysis, i.e. detecting wittiness or cheekiness, we decided on manual ranking.} as many papers as possible (i.e. as professionally permitted) in terms of the cheekiness of the title on a scale of $1-5$, where:\footnote{Not strictly, it depends on the day and the mood, or if I've seen the same acronym for the 10th time.}\\

\noindent
rank 1: completely professional title (nothing to see here, not even a modicum of effort to be witty or cheeky) \\
rank 2: there is a hint of an attempt alright \\
rank 3: fine, I smirked a little \\
rank 4: bad attempt, I rolled my eyes \\
rank 5: this is \textit{so good} (either leading to maximum eye roll and head shaking OR clapping my hands like a seal - cringe or lit, there is no in between) \\

In total, we ranked 6000 papers from $2021-2023$\footnote{Our average pace is $\sim$10 min per 100 papers, so you can do the math of how many working hours we have invested on this.}. Believe us, we would have ranked more if we could. 

\section{Results}
\label{sec:3-results}
\subsection{(An attempt) Towards a homogenized ranking system}
When two distinctly unique humans rank titles manually, it is only natural that there are some differences in ranking. This is especially important when JT is more of a zoomer\footnote{Unavoidable given that she went to college where the first FB meme group (UCBMFET) was born, and had spent considerable time curating and creating memes since young.} while TSS has more of a boomer humor. What JT finds witty and funny, most of the time TSS disagrees with her assessment. When TSS finds something eye-roll worthy, JT finds it a meh attempt at humor. We could say that our senses of humor are at odds with each other. Perhaps it could also be a generation gap between us, who knows for sure? ;)

Thus, to quantify how far off our rankings are from each other, we separately ranked the same 250 paper titles and compare the rankings. To our astonishment, our rankings are pretty similar, in particular for the titles that were ranked 2 to 4. JT ranked more titles as 1 and fewer titles as 5, and vice versa for TSS. This tells us that it is relatively easier to trigger or impress TSS compared to JT.

\begin{figure}
\centering
\includegraphics[width=0.95\columnwidth]{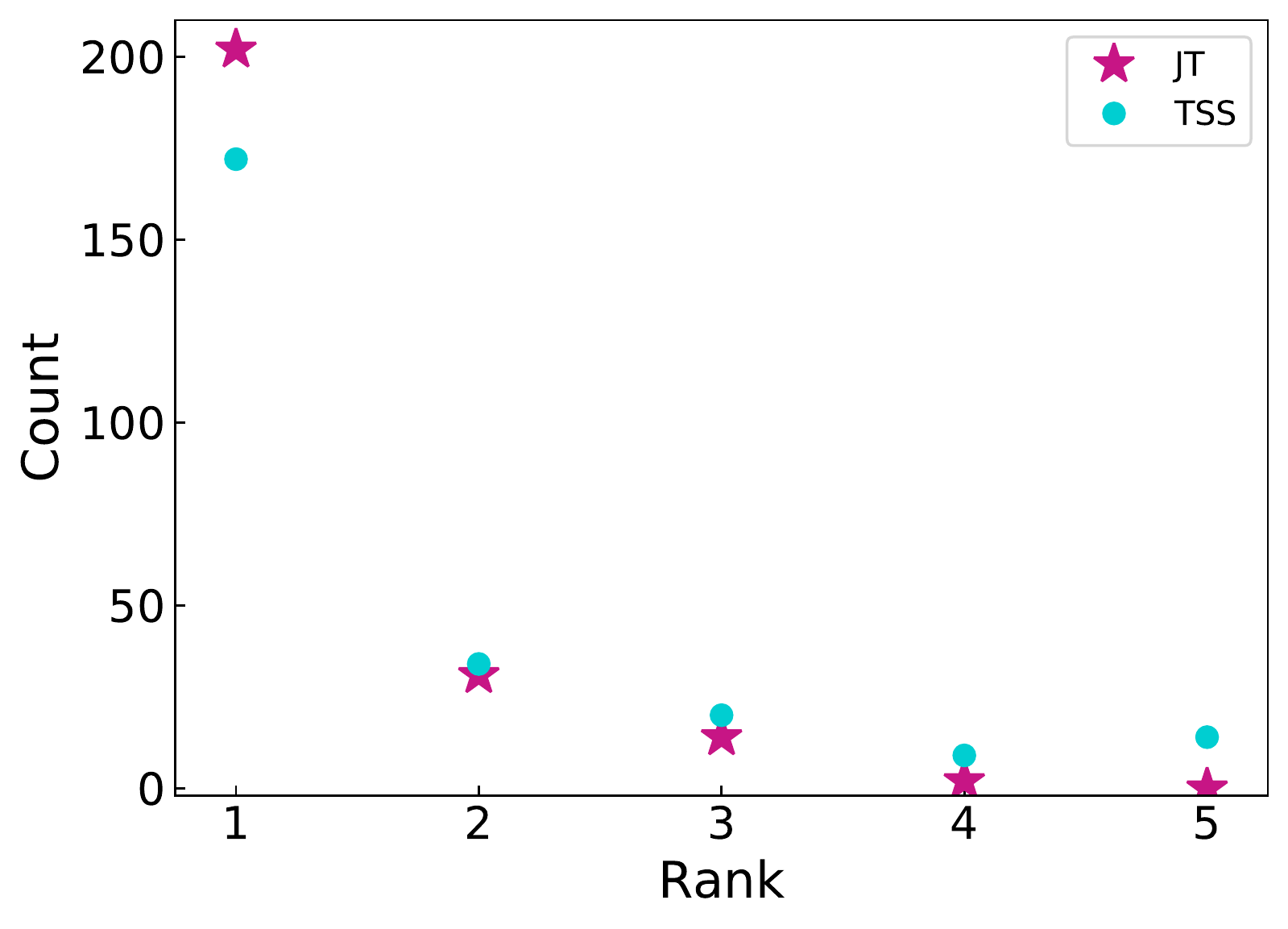}
\caption{The distribution of ranks for the same 250 paper titles that JT and TSS ranked separately. We find that our rankings are pretty consistent with each other, especially for ranks $2-4$.}
\label{fig2}
\end{figure}

Nevertheless, we find it important to correlate our rankings. This is because the ranking of the remaining 5750 titles was performed separately in an effort to reduce time wastage. Based on the statistics from the rankings of the same 250 titles, we came up with the following equation to convert from one ranking to the other:
\begin{equation}
    R_{\rm TSS} = R_{\rm JT} + \delta_{\Bar{R}} + \Bar{\sigma} r
\end{equation}
\noindent where $R_{\rm TSS}$ is the ranking of TSS, $R_{\rm JT}$ is JT's ranking, $\delta_{\Bar{R}}$ is the difference in average rank between the two authors, $\Bar{\sigma}$ is the mean standard deviation of the two rankings, and $r$ is the correlation between $R_{\rm PD}$ and $R_{\rm JT}$. We then round off the value to the nearest integer value. The value of $r$ is determined by the strength of an association between the two rankings. 
We report that our $r = 0.618$, indicating the close association between our rankings.

For the rest of the paper, we do not attempt to convert one ranking to the other. This is because our rankings are not only relatively consistent as aforementioned, but also positively correlated as proven by the positive $r$. Also, any systematic error resulting from ranking divergence will surely be smaller than our statistical errors. 

\subsection{Increase your citations with one simple hack}
\begin{figure}
\centering
\includegraphics[width=0.95\columnwidth]{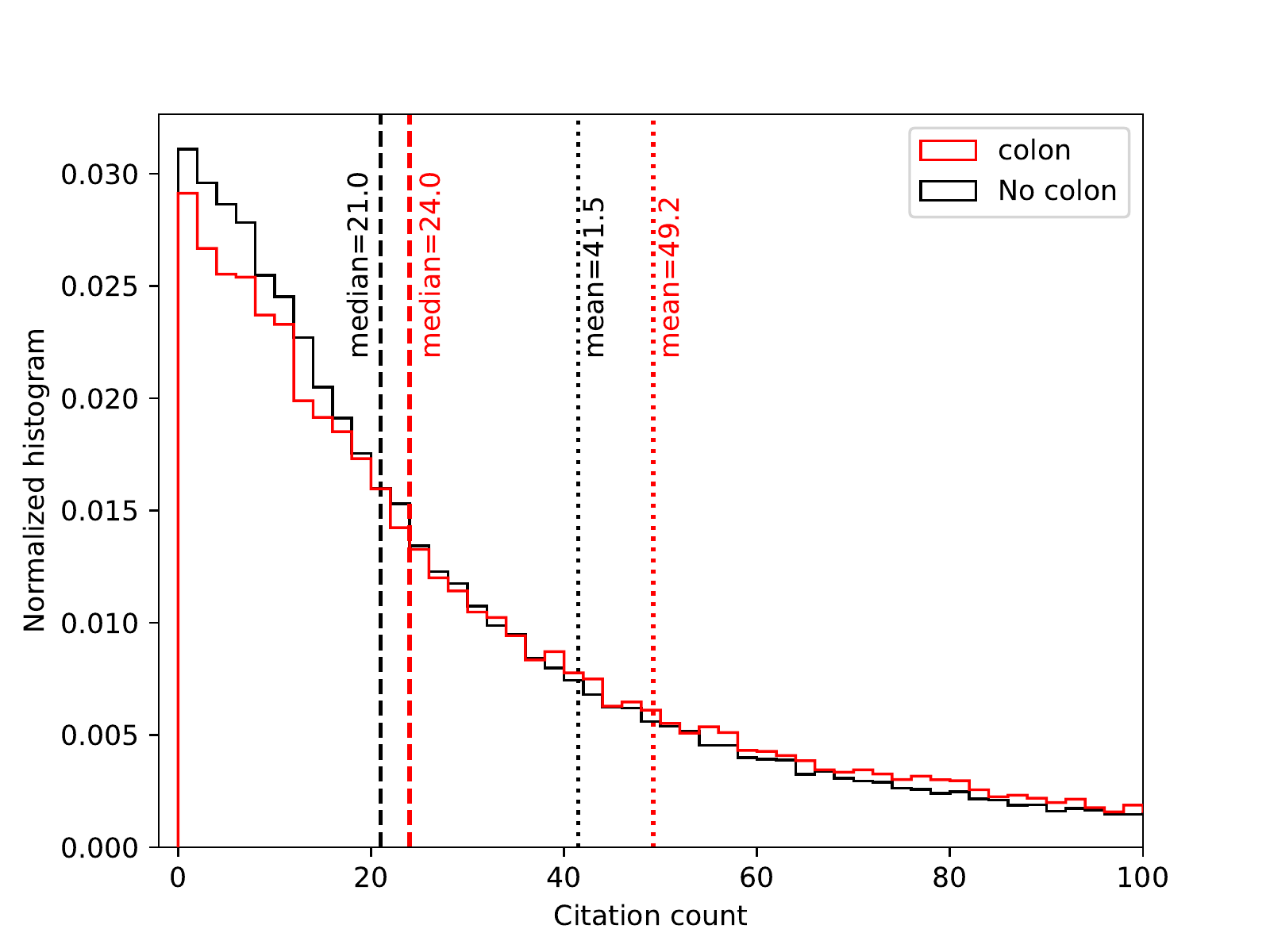}
\caption{Histogram of citation counts for papers with and without a colon in their titles. We find systematically higher citation counts for papers with colon in their titles. }
\label{fig1}
\end{figure}

Figure \ref{fig1} shows the histogram of citation counts returned by ADS for papers with and without a colon in their titles. We find that the mean and median number of citations are both higher for ``colon-titled" papers, by $\sim 15\%$! Our t- and ks-tests both confirmed that colon-titled and non-colon-titled papers represent two fundamentally different populations, where the p-values are significantly smaller than the standard threshold of 0.05. Therefore, here lies our first major conclusion --- you can increase your citation by simply including a colon in your paper title! Following this, we predict that with every colon you add, your citation rate increases by 15\%, which is to be verified in a future work\footnote{The authors shall not be held accountable if your paper gets rejected for having too many colons. }. 

\subsection{Maybe you should stay cheeky}
\begin{figure}
\centering
\includegraphics[width=0.95\columnwidth]{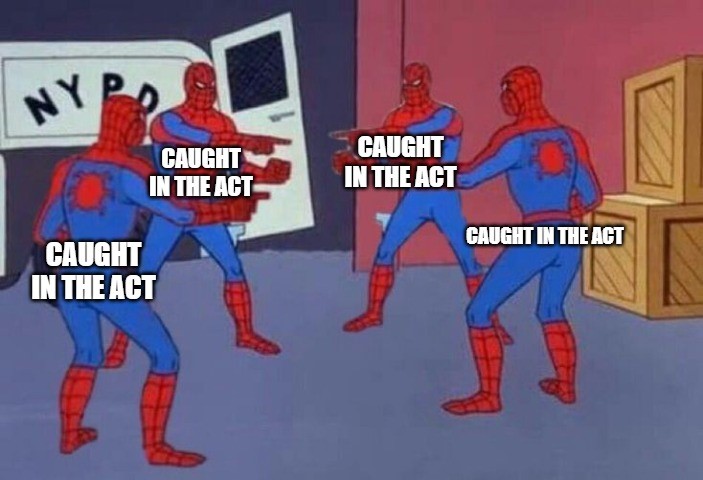}
\caption{How we imagine the IRL meetup of the 4 papers with ``Caught in the act" in their titles. These 4 papers are part of our 6000 ranked paper titles.}
\label{fig-meme}
\end{figure}

As mentioned earlier, we manually ranked 6000 papers on their cheekiness level, in which:
\begin{enumerate}[label=(\roman*)] \itemsep -4.5pt
    \item 81.40\% (4884 titles) are ranked 1,
    \item 9.83\% (590) are ranked 2,
    \item 4.17\% (250) are ranked 3,
    \item 2.65\% (159) are ranked 4, and 
    \item 1.93\% (116) are ranked 5 (see \S\ref{sec:2-methods} for the definition of rankings)\footnote{One with a keen eye would notice that these add up to 5999 instead of 6000. The missing one is due to a straggler ranked-6 paper titled ``SWEET-Cat 2.0: The Cat just got SWEETer. Higher quality spectra and precise parallaxes from Gaia eDR3''. JT just could not stop laughing her heart out upon seeing the title. It deserves the crown.}.
\end{enumerate}
We notice a few common cheeky themes, namely titles along the lines of ``One [\textit{insert noun}] to [\textit{insert verb}] them all'' (occurring once every 5.5 months), ``A tale of [\textit{insert}]'' (occurring once every 4.4 months), and ``Caught in the act'' (occurring once every 4.5 months). In Figure \ref{fig-meme}, we show what we assume would happen when the 4 papers with ``Caught in the act" in their titles meet each other IRL. To be frank, it was astonishing to JT that she caught sight of 4 of them in the 6000 papers we had ranked. It might be slightly overused at this point, may we suggest a different catchphrase to catch the reader's attention?

\begin{figure}
\centering
\includegraphics[width=0.95\columnwidth]{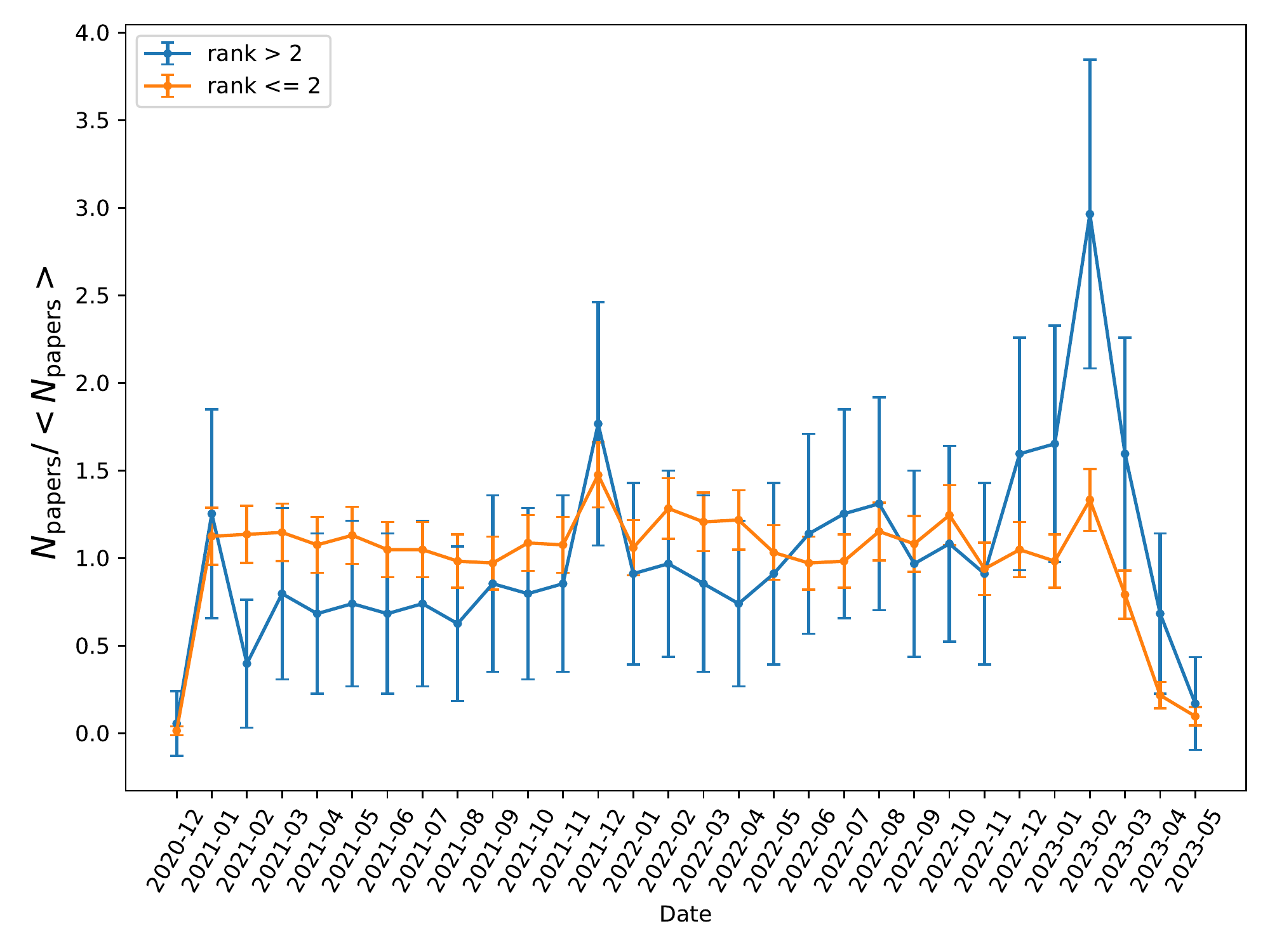}
\caption{Number of papers divided by the mean vs. time for those ranked $>$ 2 (cheeky) and those ranked lower (normal). Other than a couple of blip events on 2012-12 and 2023-02, the number of cheeky papers appears consistent with a flat trend within the error bars. }
\label{fig3}
\end{figure}

Let us now switch gears to look hard at the time evolution of the number of cheeky papers. In figure \ref{fig3}, we plot the trend with time\footnote{No, we do not live in the future. We are not sure why ADS query returns dates in the future. And the two of us are too tired to dive deep into investigating this.} for two groups of papers, those with cheeky titles (ranked $>$ 2) and those with normal titles (ranked $\leq$ 2), where the errors bars are Poisson errors \citep{Gehrels1986}. We divide the number of papers for each group by their respective means for visualization purpose, since there are significantly more normal papers than cheeky papers. Although there is a mild indication of an upward trend in the number of cheeky papers, the error bars are consistent with a flat trend sadly.

\begin{figure}
\centering
\includegraphics[width=0.95\columnwidth]{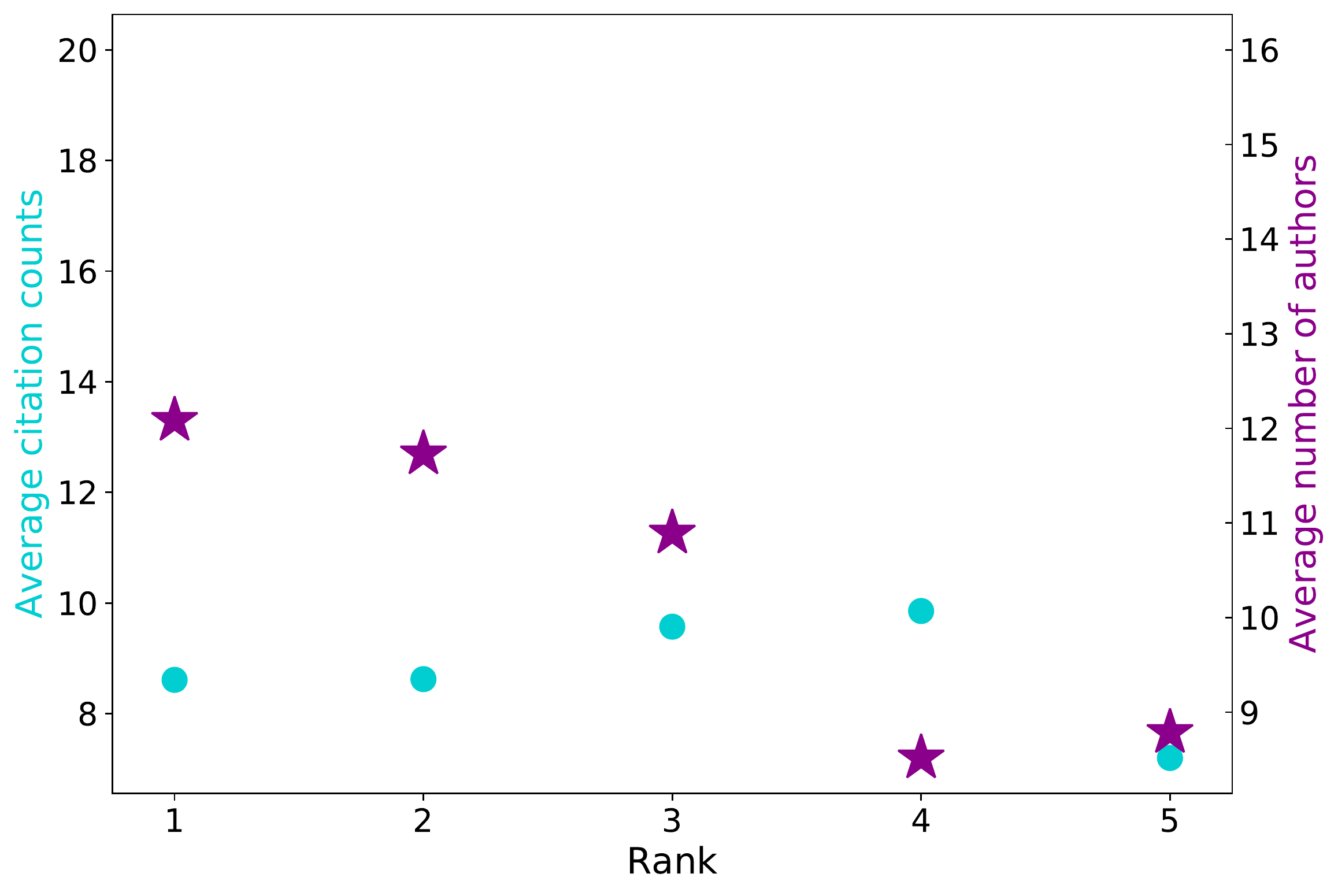}
\caption{The average citation counts of papers and the average number of authors as a function of a paper title's rank. We see that paper titles ranked higher in cheekiness have a lower average number of authors. The average number of citations also increases with the level of cheekiness, but drastically drops when the cheekiness gets too high.}
\label{fig4}
\end{figure}

Finally, we study the correlation between the cheekiness ranking with citation counts (does cheeky paper get more cites?) and number of authors (does cheeky paper come from smaller or larger groups?). The results are shown in Figure \ref{fig4}. First and foremost, we note that the cheekiness of a paper is strongly anti-correlated with the number of authors, with cheekier papers having on average $<10$ authors. We surmise this could be due to first authors having more autonomy (less group politics and conflicting co-authors feedback) when residing in smaller groups. In addition, we find that the average citation tends to increase with cheekiness (much to the chagrin of one of the authors). Nevertheless, it appears that being too cheeky can be off-putting, as we observe a drastic drop in citation for papers that are ranked 5. 

\section{Conclusion}
\label{sec5: conc}
In summary, we study the correlation between paper titles and their number of citations, and found strong indications that cheeky titles appear to make a difference after all (LOL).  We present a reliable ranking scheme based on gut reaction (of which first impressions are based) on how to decide where your paper lies on the cheekiness scale (see \S\ref{sec:2-methods}). 

Putting all that we have learned together, overall we suggest researchers aim for a level 4 cheeky title to achieve maximum bang for your buck. Make sure to include a colon for compounded effects. We also recommend researchers work in smaller groups and/or restrict their author list to $<10$ people to guarantee maximum autonomy in naming their papers. Especially for you, we also provide a prime example in this paper, as illustrated in our title. With these, may your paper(s) live long and prosper. Prayge.

\section{Acknowledgement}
\noindent The authors acknowledge the time and effort researchers invested in executing original research, presenting methods and findings in papers, as well as embellishing the papers with eye-catching titles. They are not the easiest of tasks. We also thank the entertaining and/or eyeroll-inducing moments just from reading the paper titles -- it was a nice change of pace in our mundane life. We thank you for your creativity :) We also acknowledge Matplotlib \citep{Hunter:2007}, Numpy \citep{harris2020array}, and Scipy \citep{2020SciPy-NMeth}.

\bibliography{biblio}{}
\bibliographystyle{aasjournal}

\end{CJK*}
\end{document}